\documentclass[fleqn,10pt]{wlscirep}

\usepackage{graphicx}
\usepackage{dcolumn}
\usepackage{bm}
\usepackage{multirow,caption}
\usepackage{blindtext}
\usepackage{mathtools}
\usepackage[utf8]{inputenc}
\usepackage[T1]{fontenc}
\usepackage{amsmath}
\usepackage{adjustbox}
\usepackage{amsmath}
\usepackage{hhline}
\usepackage{color}
\usepackage{xcolor}
\usepackage{float}
\usepackage{pbox}
\usepackage[normalem]{ulem}
\usepackage{array}
\usepackage{caption}
\usepackage{subcaption}
\usepackage[thinlines]{easytable}
\usepackage{ulem}
\usepackage{booktabs}
\usepackage{graphicx}
\usepackage{hyperref}

\title{ Characterizing functional brain networks and emotional centers based on Rasa theory of Indian aesthetics}

\author[1]{Richa Tripathi}
\author[2, 1]{ Dyutiman Mukhopadhyay}
\author[1]{Chakresh Kumar Singh}
\author[1]{Krishna Prasad Miyapuram}
\author[1,*]{Shivakumar Jolad}

\affil[1]{Indian Institute of Technology Gandhinagar, Gandhinagar- 382355, India}
\affil[2]{University College London, WC1E 6BT, UK} 

\affil[*]{shiva.jolad@iitgn.ac.in}

\begin{abstract}
In Indian history of arts, \textit{Rasas}  are the aesthetics associated with any auditory, visual, literary or musical piece of art that evokes highly orchestrated emotional states.
In this work,  we study the functional response of the brain to movie clippings meant to evoke the \textit{Rasas} through network analysis. We extract functional
brain networks using coherence measures on EEG recordings of film clips from popular Indian Bollywood movies representing  nine \textit{Rasas} in the Indian \textit{Natyasastra}. Structural and functional network measures were computed for these brain networks, averaging over a range of significant 
edge weights, in different brainwave frequency bands. We identify  segregation of neuronal wiring in the brain into modules using a community detection 
algorithm. Further, using mutual information measure, we  compare and contrast the modular organizations of brain network 
corresponding to different \textit{Rasas}. Hubs identified using centrality measure reveal the network nodes that are central to information propagation across all \textit{Rasas}. We also observe that the functional connectivity is suppressed when high-frequency waves such as beta and gamma are dominant in the brain. The significant links causing differences between the \textit{Rasa} pairs are extracted statistically.
\end{abstract}

\begin{document}

\flushbottom
\maketitle

\thispagestyle{empty}

\section*{Introduction}

Emotions affect the body functioning in multiple ways such as by introducing electro-physiological changes in the facial muscles, electrical activity in the brain, changes in heartbeat rates and respiration \cite{izard1977emotions}. These physiological changes involve all the Sub-Systems of the body to a small or large extent and hence affect our perception, thoughts and actions. Emotion states also have a significant effect on the social development of the individuals\cite{izard2013human}. Apart from the cognitive origin, at the neurological level, it is believed that changes in the density of neural stimulations \cite{tomkins1962affect} drive the emotions through agents such as inborn releaser and external audio and visual stimulations. The functional changes that induce emotions through neural segregation of emotional and non-emotional pathways can be inferred using brain
imaging and EEG studies of individuals \cite{pfurtscheller1999event}.

In the recent years, analysis of brain networks constructed from EEG and advanced brain imaging techniques has contributed to the understanding of the complex structure  and functioning of the human brain \cite{bullmore2009complex, sporns2004organization} and has  given clues to understanding its higher
cognitive exhibits such as emotions and reasoning abilities \cite{lee2014classifying,pessoa2008relationship,mcmenamin2014network}. The functional 
brain networks studies have given reasonable evidence against the locationist approach of functional segregation of distinct brain regions corresponding to specific emotion categories in favour of constructionist approach i.e. different emotions involve brain circuits comprising of brain areas not specific to a particular emotion \cite{lindquist2012brain, pessoa2008relationship, pessoa2010emotion}. Emotional stimuli affect large scale functional brain networks which can be evaluated in terms of parameters such as node betweenness and network efficiency \cite{pessoa2017dynamic}. 
The human brain can be decomposed into multiple, distinct, and interacting networks such as salience network, executive control network and task negative network, 
and emotional stimuli can differentially affects these subnetworks \cite{menon2015brain,pessoa2017dynamic,mcmenamin2014network}.
For example, strong changes in salience networks have been reported during watching an aversive movie segment as compared to watching a neutral movie segment through fMRI
studies\cite{hermans2014dynamic} . 

Studies on emotion based on the analysis of single-electrode level based EEG in the frequency domain have demonstrated the association of emotion with asymmetric activity in the frontal brain in alpha band. Ekman and Davidson (1993) \cite{ekman1993voluntary} in their study found that voluntary facial expressions of smiles of enjoyment produced high left frontal brain activity. On the contrary, Coan \textit{et al.} \cite{coan2001voluntary} reported that face
expressions of fear produce low frontal activation. Also, in a study by Sammler and colleagues \cite{sammler2007music}, it was proposed that increase in frontal
midline theta power was associated with happy emotion. Apart from using alpha and theta asymmetry for emotional classification, 
power spectral features of EEG signals were extracted in five frequency bands and used for emotion recognition using Deep Learning network architecture with considerable 
accuracy{\cite{jirayucharoensak2014eeg}. It has also been demonstrated that different patterns of functional connectivity are associated with different emotional states in either single or combined frequency bands ascertaining distinct response patterns of the central nervous system to different emotional stimuli \cite{lee2014classifying}.
 
In previous studies on classifying emotions, researchers have largely focused only on few contrasting dimensions such as threat-safe, sadness-happiness, positive-negative-neutral and with a small number of recording sites of EEG activity\cite{nie2011eeg, dmochowski2012correlated, lee2014classifying}. Also, the film-based studies on emotions and functional response of 
brain has been previously carried out primarily based on the Western classification of emotions which are happiness, sadness, fear,  anger,  surprise, disgust
\cite{ekman1984expression, ekman1992argument}. However, audio-visual stimuli such as film viewing are capable of evoking sentiments which might be due to interplay the basic emotions (dominant states) and transitory and temperamental states of emotions. Indian
\textit{Rasa} theory, as described in the \textit{Natyasastra} has outlined these sentiments or \textit{Rasa}s as the superposition of these states of emotions \cite{ghosh1951natyasastra}.

In this work, we study the functional response of brain captured through 128 channel EEG recordings to movie clippings eliciting the response corresponding to the nine \textit{Rasas} classified according to Indian \textit{Natyasastra}. We have recorded the response of participants to a wide range of audio-visual stimuli from clippings of popular Indian Bollywood movies in Hindi. We have analysed the patterns of brain activity in different frequency bands corresponding to Brainwaves \cite{tatum2014ellen} and examined whether these activities would show specific neural signatures across viewers.   Different network measures were used to characterises the structural and functional differences among the \textit{Rasa} states.  Community structure corresponding to different \textit{Rasa}s were computed to highlight the specificity to different brain areas or circuits and fill the gap of brain network analysis.  Modular organisations of networks corresponding in each of the frequency bands were compared using information flow measure to quantify the difference between the \textit{Rasa}s. Most central of the nodes were identified using leverage centrality measure for each of the networks, enabling us to find the most significant nodes for emotion perception and segregate designated areas for particular emotion processing. The significant links across all \textit{Rasa}s pairs were identified statistically to locate origins of differences in functional networks using Honest Significant difference test across emotion groups after the F-test as followed in a work by Lee (2014)\cite{lee2014classifying}.

\section*{Background}

\subsection*{\textit{Rasa} and \textit{Natyasastra}}

A major source of the Indian system of classification of emotional states comes from the ‘\textit{Natyasastra}’, the ancient Indian treatise on the performing arts, 
which dates back to 2nd Century AD (pg. LXXXVI:  \cite{ghosh1951natyasastra}). The ‘\textit{Natyasastra}’ speaks about ‘sentiments’ or ‘\textit{Rasas}’ 
(pg.102: \cite{ghosh1951natyasastra}) which are produced when certain ‘dominant states’ (\textit{sthayi bhava}), ‘transitory states’ (\textit{vyabhicari bhava}) and 
‘temperamental states’ (\textit{sattvika bhava}) of emotions come together (pgs.102, 105:  \cite{ghosh1951natyasastra}).  This \textit{Rasa} theory, which is still widely 
followed in classical Indian performing arts, classifies eight \textit{Rasas} or sentiments which are: \textit{Sringara} (erotic), \textit{Hasya} (comic), \textit{Karuna} (pathetic), 
\textit{Raudra} (furious), \textit{Veera} (heroic), \textit{Bhayanaka} (terrible), \textit{Bibhatsa} (odious) and \textit{Adbhuta} (marvelous). There was a later addition of the ninth sentiment or \textit{Rasa} called \textit{Santa} (peace) in later Sanskrit poetics (pg.102:  \cite{ghosh1951natyasastra}). 

The eight dominant states (\textit{sthayi bhava}) of emotions which give rise to these corresponding eight \textit{Rasas}
or sentiments are: love (for \textit{Sringara}), mirth (for \textit{Hasya}), sorrow (for \textit{Karuna}), anger (for \textit{Raudra}), energy (for \textit{Veera}), terror (for \textit{Bhayanaka}), disgust 
(for Bibhatsa) and astonishment (for \textit{Adbhuta}). 
Apart from these there are thirty three ‘transitory states’ (\textit{vyabhicari bhava)} and eight ‘temperamental states’ (\textit{sattvika bhava}) of emotions mentioned in the 
Natyasastra which play a critical role in the generation of \textit{Rasas} or sentiments (pg.102: \cite{ghosh1951natyasastra}).

\subsection*{ Western emotional classification and \textit{Rasa} theory}

According to Western version of emotion classification by Ekman, there are basic or universal emotions  \cite{ekman1971constants, ekman1984expression},
\cite{ekman1992argument} which are happiness, sadness, fear,  anger,  surprise, disgust. 
However, there are also background emotion sets which are:
wellbeing-malaise; calm-tense; pain-pleasure; \cite{damasio1999feeling} as well as self-referential social emotions which are: embarrassment, guilt, shame, 
jealousy, envy, empathy, pride, admiration. \cite{oatley1987towards, bennett1991role,  diener1999subjective, hareli2006role, leary2000nature,  
	tangney1995self}. Also, there are pioneering works of scientists like Lisa Feldman Barrett \cite{barrett2006emotions} who questions Ekman's concepts
of discreteness of emotions. 
Apart from defining the boundaries of universal emotions, research in emotion science also places equal emphasis in trying to understand the interplay of the 
different orchestrated processes that give rise to a basic emotion \cite{barrett2006emotions, panksepp1994effects, kagan1997temperament}.
For example, Panksepp \cite{panksepp1994effects , panksepp2010affective} defined primary processes of emotion (not the basic or universal emotions of Ekman) as primary sub-neocortical processes of emotion having their corresponding affective states which can be artificially generated by brain stimulation in animals. Also, the four-step cascade process proposed by Jerome Kagan \cite{kagan1997temperament}
defines a provocative event which leads to brain-change and subsequently leads to a feeling and the interpretation of the feeling gives rise to an emotion. 

We can find startling similarities between the \textit{Rasa} theory (its concepts of the generation of \textit{Rasa}s from the Bhavas) with the works of 
Panksepp and Kagan \cite{panksepp1994effects, panksepp2010affective, kagan1997temperament,parrott2001emotions}.  However, there had been very little previous work done on the perception
or brain science of emotional states based on \textit{Rasa} theory mostly due to the lack of awareness regarding the science of the \textit{Rasa} theory among the scientific community. One behavioural study was conducted by Hejmadi \cite{hejmadi2000exploring}, which investigated the identification of these emotions across cultures. An image processing study was conducted in Ref. \cite{srimani2012analysis} for investigating the variations in facial features based on nine \textit{Rasas}. The study proposes a tool for the design of intelligent
emotion recognition system but does not offer a psychophysical or brain-based perspective on how these individual emotions differ in the way they affect our perceptual process. There have
been no brain-imaging based works on \textit{Rasa}-theory comparable with similar works done on emotions based on Western literature. We analyze through EEG, the patterns of brain activity generated through nine different emotional responses based on Indian \textit{Rasa} theory while watching films and to find out whether these activities would show specific neural signatures across viewers.

\section*{Results}

\subsection*{Brain Network Characterization}

Functional properties of complex networks are largely determined by the statistical properties of its structure. We have calculated six well-known 
network measures- Clustering Coefficient ($CC$), Characteristic Path Length ($L$), Network Density ($d$), Local and Global efficiency ($LE, GE$), and 
Small-worldness index ($SW$) using Area Under Curve method (see Methods for details).  

The clustering coefficient ($CC$) is a measure of network segregation. The high global clustering coefficient of the brain network quantifies functional 
segregation in the brain for specialized processing to occur within densely interconnected groups of brain regions. The characteristic path length $L$ 
(average shortest path length between all pairs of nodes in the network - see e.g. \cite{watts1998collective}) on the other hand, is the most commonly used 
measure of functional integration \cite{rubinov2010complex}. Network density $d$ measures the number of links present as compared to the number of possible 
connections in the network. Global efficiency of network quantifies information exchange across the whole network where the information flow occurs on a 
global scale. The local efficiency quantifies the network's robustness to failure in the local node neighbourhood i.e. measure of information flow between 
the neighbours of a node if it is removed from the network.

Small-worldness index ($SW=\frac{C/C_{rand}}{L /L_{rand}}$ , see Methods) measures the amount of deviation of actual networks from the random networks \cite{watts1998collective}. Functional brain networks with high correlations retained show many biologically relevant features such as "small-world" behavior\cite{watts1998collective,bassett2006small, hayasaka2010comparison}, existence of hubs and high network modularity. Small-world topology supports distributed as well as segregated information processing at the same time and is highly economical in terms of wiring cost involved in this information processing \cite{bassett2006small}.

In Fig.~\ref{fig:NetworkMeasures}, we plot the six measures described above for all the \textit{Rasas} across all frequency bands.  Network density $d>0.45$ for all the networks obtained through AUC method shows that they are dense. Average shortest path length lies between 1.2 to 1.8, indicating the ease of information propagation.  Network density decreases with increasing frequency, except in the gamma band- where it becomes denser. Correspondingly we see a drop in path length. Clustering coefficient shows small variations from 0.75 to 0.9. Gamma band has greater spread than all other bands.  \textit{Veera } and \textit{Bhayanaka} shows the maximum difference across $\delta, \theta, \alpha$, and $\beta $ bands.  SW greater than one for all \textit{Rasas},  indicating deviation from absolute randomness as well as regularity of connections in the brain.  Global efficiency of the networks is high - ranging from 0.7 to 0.87.  Local efficiency is even higher- 0.88 to 0.95, indicating high robustness to the failure of nodes.  The trend (increase or decrease of network measures with frequency) is flipped at high frequency in the $\gamma$ band. This is seen for all the \textit{Rasas} and for each network metric, suggesting a dynamic reorganisation of brain structure at high frequency. Information flow at gamma frequency the 
brain has a special structural organisation quite very different from the one at just lower frequency band.

Across all the frequency bands, except $\gamma$ it is seen that the \textit{Rasas} \textit{Bhayanaka} and \textit{Veera} are at opposite extremities in terms of the values of
Network measures. The order of \textit{Rasas} across the extent of network properties enveloped by properties of these two \textit{Rasas} is \textit{Bhayanaka}, \textit{Karuna}, \textit{Raudra},
\textit{Santa}, \textit{Hasya}, \textit{Sringara}, \textit{Bibhatsa}, \textit{Adbhuta} and \textit{Veera}. This order is not followed in gamma bands, where  \textit{Karuna} and \textit{Hasya} are at the ends of
\textit{Rasas} spectrum. 

\begin{figure*}[t]
	\includegraphics[scale=0.9]{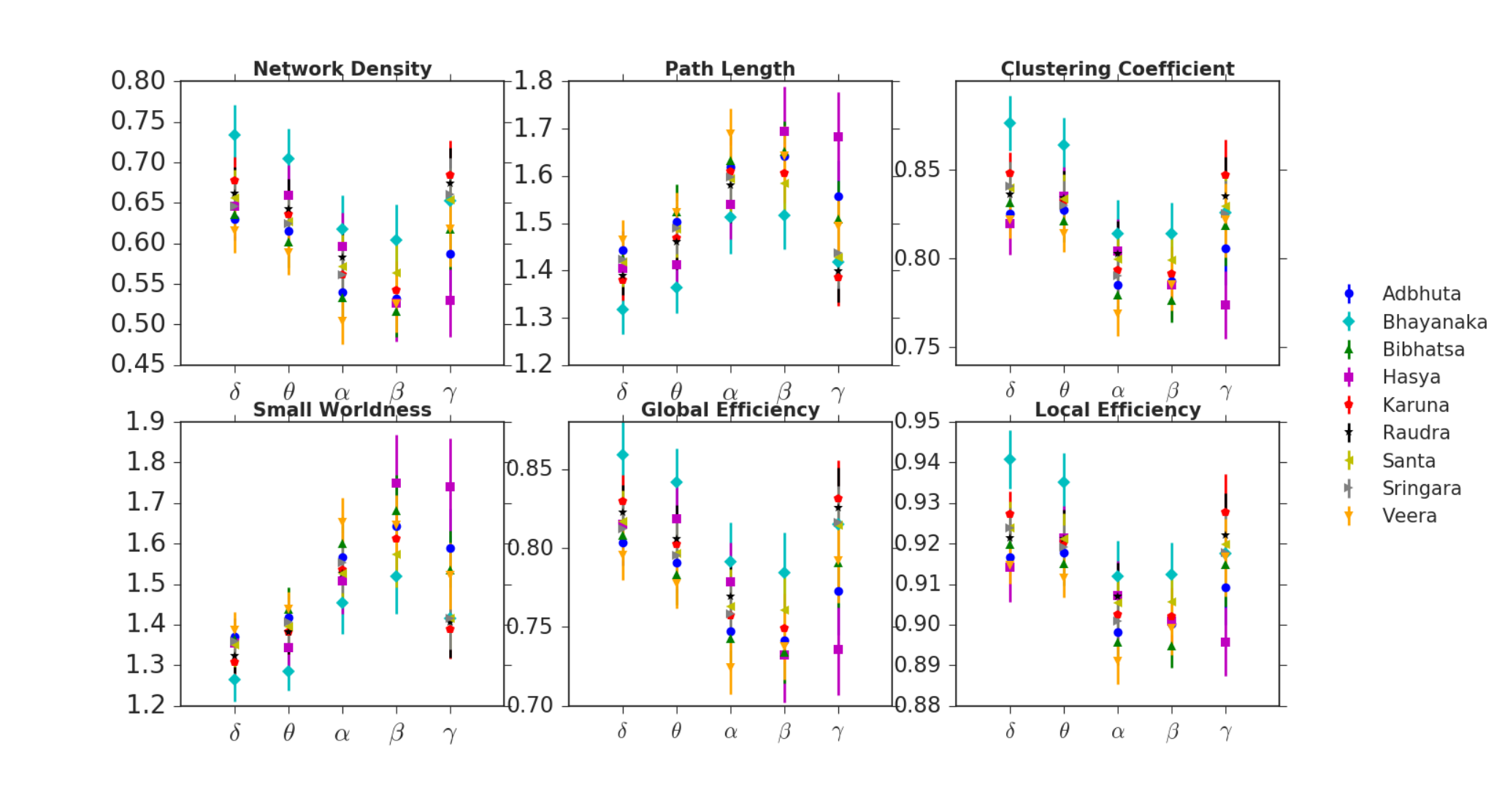}
	\caption{Network measures Network Density (d), Path Length (L), Clustering Coefficient (CC) , Small Worldness , Global Efficiency (GE), and 
	Local Efficiency (LE) (see text for details) of brain networks corresponding to different \textit{Rasas} across the five frequency bands
	($\delta, \theta, \alpha, \beta$ and $\ \gamma$) along with their error bars. }
	\label{fig:NetworkMeasures}
\end{figure*}

\newcommand{\squeezeup}{\vspace{-2.5mm}}

\subsection*{Edge Weight distribution}
Edge weights denote the strength of connections between the nodes.  The cumulative edge weight distribution of all \textit{Rasas } is shown in supplementary information Fig.SI2.  We find that edge weights are considerably high for $\delta$, $\theta$ and $\alpha$ bands for all networks and lower for $\alpha$ and $\beta$ bands. Through which we infer that functional connectivity of brain for emotion perception is stronger in $\delta$, $\theta$, and $\gamma$ bands whereas the emotional pathways have lesser average information flow in $\alpha$ and $\beta$ frequency regimes. We also observe that  \textit{Raudra}, \textit{Bhayanaka} and \textit{Hasya} have higher correlations specially in $\delta$, $\theta$ an $\alpha$ bands as compared to other \textit{Rasas}.

\subsection*{Community structure}
Quantitative analysis of brain using complex networks measures has revealed the presence of highly connected hubs and significant modular architecture\cite{achard2006resilient}, apart from showing small-worldness.  Modules are functionally specialised and spatially localised groups of nodes that function together in unison to integrate the information globally that they process locally. 

In Fig.~\ref{fig:com}, we show community structure of brain networks corresponding to all the \textit{Rasas}, with top 10$\%$ of the edges retained,  extracted using the modularity optimisation algorithm in Gephi. The communities were then extracted from the complete networks (with all the edges retained, after removing the 5 $\%$ weakest connections) corresponding to all the \textit{Rasas}.
All the networks were highly modular with modularity, $Q \geq 0.58$, and had with five to six communities in each case. Broadly, the community structure appears  similar across all networks, with one major community in the frontal lobes (C1), two in the left and right hemisphere's central/parietal brain
regions (C2, C3), one that encompasses a large area in the visual cortex in the occipital areas (C4) and two other smaller communities along the left and right temporal regions (C5, C6). The  \textit{Bibhatsa} \textit{Rasa} shows split of the community in the parietal lobe. The communities C4 and C5 are merged into one for most of the graphs. However, the frequency band wise networks as shown in the supplementary information (Fig. SI 3), generally shows them as separate communities.

\begin{figure*}[!h]
	   \centering  
	\includegraphics[scale=0.9]{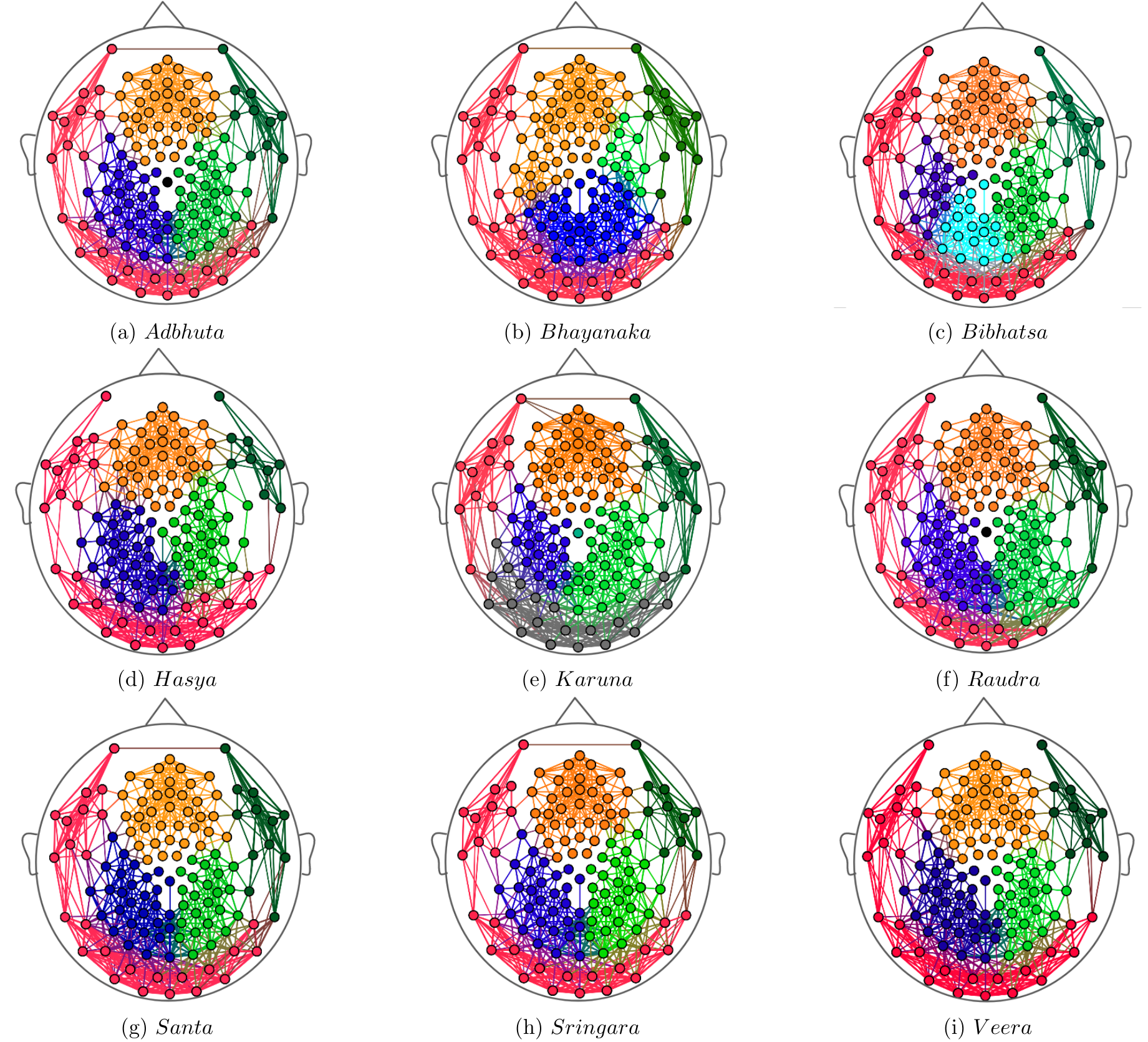}
\caption{Brain Networks for various \textit{Rasas} with 10 per cent highest edge weight edges, organised into clear community structure. For all the \textit{Rasas} the network is dissociated into 5 to 6 communities. One can notice the changes in the community structure from this figure, while there are a few clear communities consistently present across all the networks. The modularity value for network partition was above 0.5 for all the \textit{Rasas}, depicting highly modular structure at the level of strongest connections in the network. }
\label{fig:com}
\end{figure*}

\begin{figure*}[!h]
	\centering
	\includegraphics[scale=0.9]{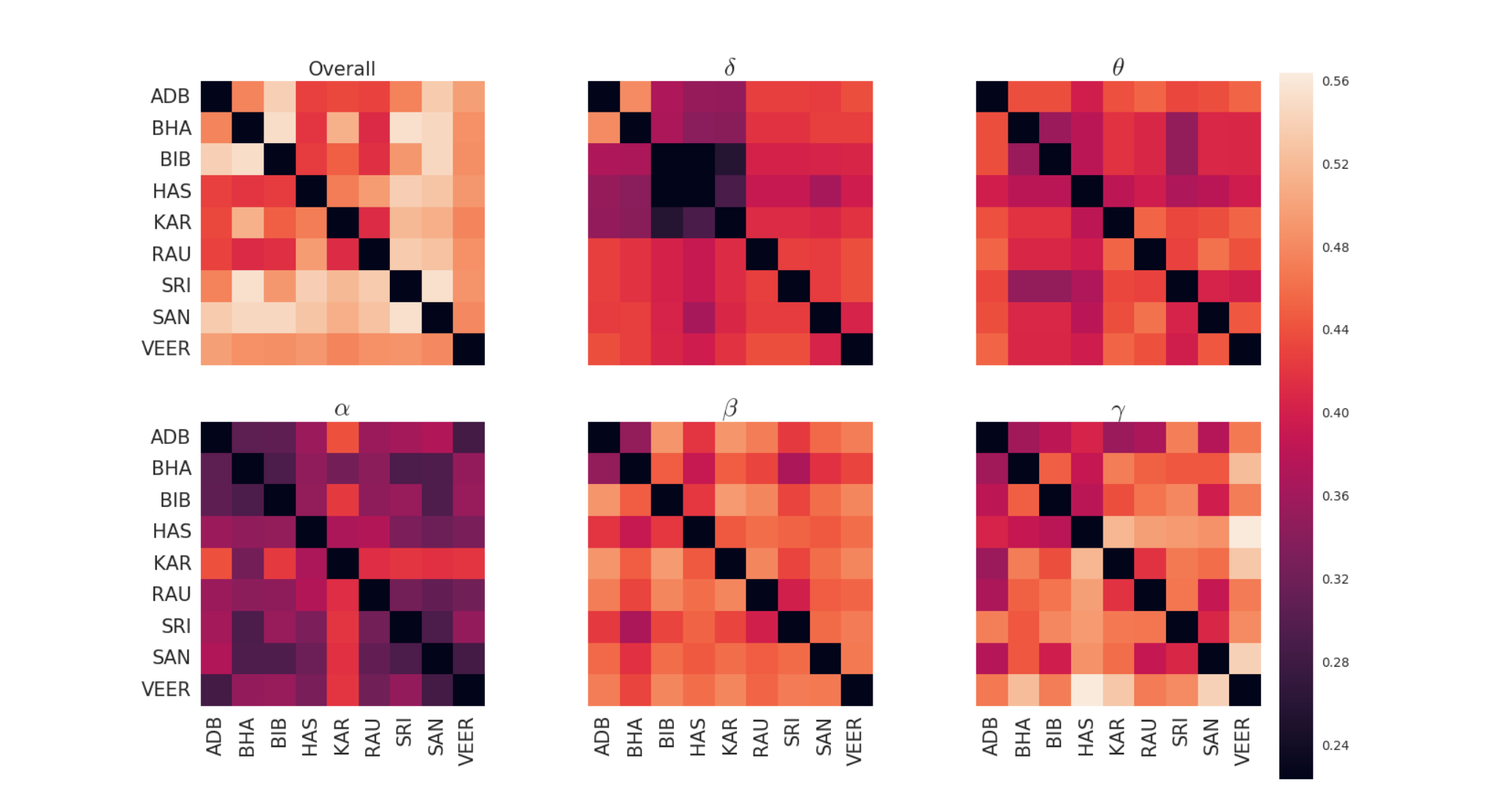}
	\caption{Variation of Information (Eq. \ref{Eq:VI}) matrices for various \textit{Rasas} in different frequency bands arranged in order \textbf{overall, $\delta$, $\theta$, 
			$\alpha$, $\beta$ and $\gamma$}, from left to right and top to bottom. It can be seen that \textit{Rasas} have more similar modular 
			organization in $\delta$ and $\alpha$ bands than in $\theta$, $\beta$ and $\gamma$ bands.}
	\label{fig:VI}
\end{figure*}

\subsection*{Distance between Networks}

To compare the similarities and differences in the network’s modular organization (communities), we use the Variation of information (VI) 
measure \cite{karrer2008robustness}  
(See Eq.~\ref{Eq:VI} in Methods).  The VI measure signifies the difference in the modular organisation of the networks, summarising the differences in segregation and network wiring. In Fig.~\ref{fig:VI}, we show the Variation of Information measure for different \textit{Rasas} in different frequency bands. Networks that are farthest from each other are shown in brighter cell colour ($VI \simeq 0.5$), and those which are closest are shaded dark ($VI \simeq 0.0$). Six VI measure matrices were obtained - one corresponding to overall spectrum and five other corresponding to each of frequency bands (see Fig.~\ref{fig:VI}).  
We have evaluated the distance matrix between all \textit{Rasa}s based on the total power to visualise the overall differences in network partitions. In the supplementary information (Fig. SI 1), we have compared the modular organization of each emotion across all frequencies i.e with one emotion at a time. 

For all the pairs of \textit{Rasas}, the VI falls in the range 0.22 to 0.56. The colour bar (common for all matrices) range is chosen to represent the maximum variation in VI. 
We observe that, for the full network, the dynamic range of VI variation is much small, hence we cannot infer much about the similarity or dissimilarity of different network organizations. In the lowest frequency $\delta$ band -  \textit{Bibhatsa}, \textit{Hasya} and \textit{Karuna} are closest to each other. Maximum difference is between  \textit{Adbhuta} and \textit{Bhayanka}. For $\theta$  band- \textit{Bhayanaka}, \textit{Bibhatsa} , \textit{Sringara} are relatively closest. Others are far from each other. The $\alpha$ band is the most distinctive frequency band where all the \textit{Rasas} are much closer to each other (except the \textit{Karuna}). The $\alpha$ band activity for all the \textit{Rasa}s , except \textit{Karuna}  seems indistinguishable from each other.
For $\beta$ band, the pairs \textit{Adbhuta}, \textit{Bhayanaka},  \textit{Raudra}, and \textit{Sringara} are the closer to each other, and  \textit{Hasya} and \textit{Adbhuta} are the farthest. In the $\gamma$ band, \textit{Veera} is farthest from \textit{Santa}, \textit{Hasya} and \textit{Bhayanaka}. \textit{Adbhuta} is much closer to \textit{Bhayanaka} and \textit{Karuna}.

 \begin{figure*}[!h]
 	\centering
 	\includegraphics[scale=0.9]{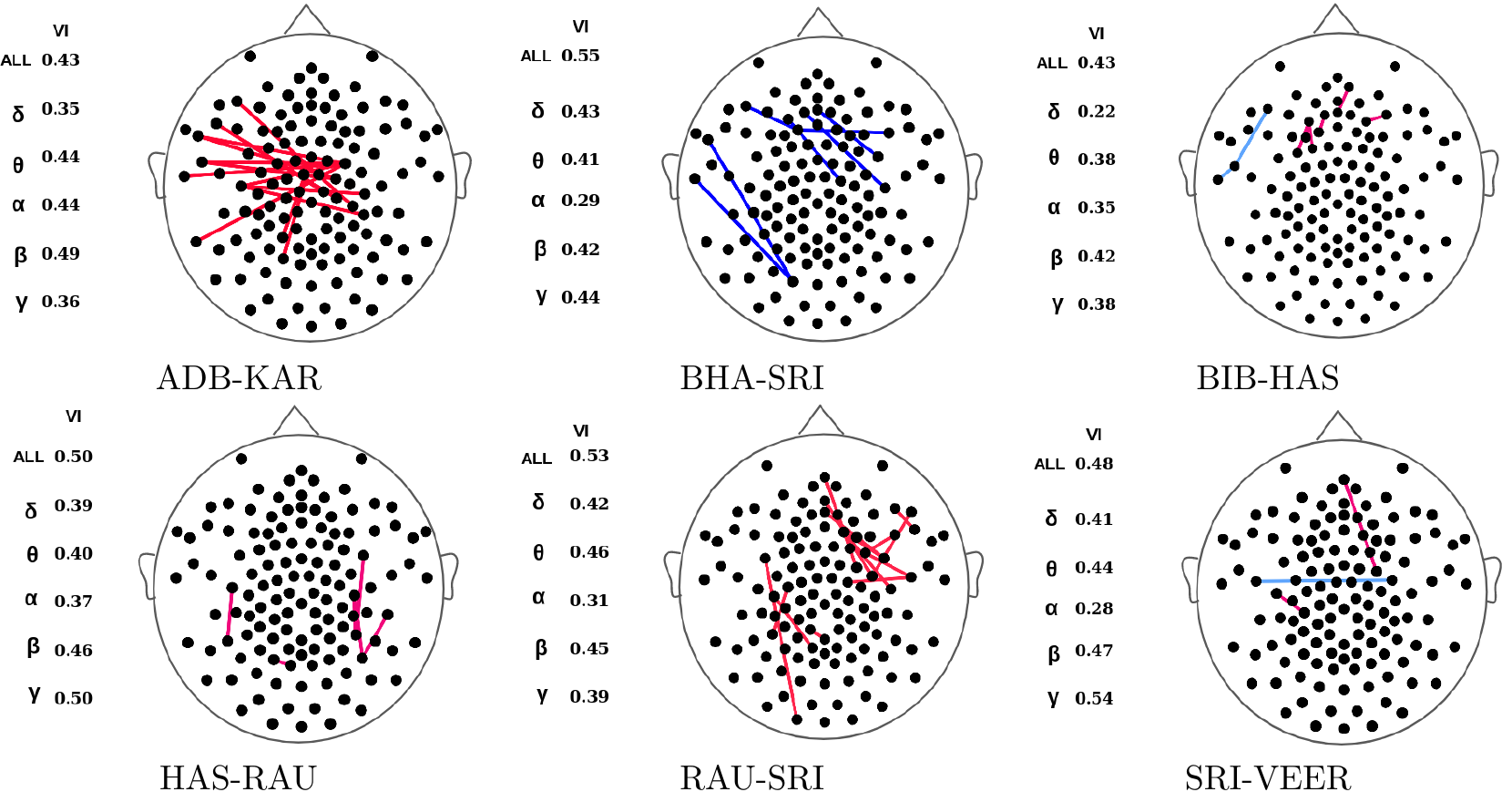}
 	\caption{Brain maps showing Honest Significant Differences of coherences for groups, (a)Adb-Kar (b)Bha-Sri (c)Bib-Has (d)Has-Rau (e)Rau-Sri and 
 	(f) Sri-Veer, with higher (Blue) and lower (Red) coherences for the group on the left. The values in the panel on the left of each figure is the Variation of information
 		measure corresponding to the two \textit{Rasas} in all frequency bands and in overall spectrum.   }
 	\label{fig: HSD}
 \end{figure*}

\subsection*{Honest significant edges among different \textit{Rasas}}

To assess whether a particular edge weight in the brain network of a given \textit{Rasa} is on average higher, or lower compared to other groups, we performed F-test on all the edge weights against the null hypothesis that all \textit{Rasas} (groups) yield the same mean edge weight. The edge weights that pass the F-test were the ones considered significant across all groups (with significance $\alpha \leq 0.05$). But, this does not reveal
which individual group means contribute to this significance. 
To identify where the group differences lie, Tukey’s Honest Significant Difference (HSD) was performed as a post-hoc analysis on the edge weights that were
proven significant by the F-test. Tukey's test enables us to determine two \textit{Rasa}s/ groups, corresponding to each of these significant edge weights.

We extracted significant edges for all 36  pairs of un-thresholded networks.  The pairs are shown in the Fig.~\ref{fig: HSD} shows significant edges for six such pairs. The edges represent higher (Blue) and lower (Red) coherence for the group on the left in each pair. The group means of edge weights that contribute to the significant difference is spread over the entire space for some and it is highly localized for others.Between \textit{Adbhutam} and \textit{Karunyam} , edges in the central and  parietal left regions show significant positive 
differences. \textit{Bhayanka} and \textit{Sringara}  have differences in the frontal and temporal left regions. \textit{Raudra} and \textit{Sringara} has differences in 
right frontal and left parietal regions. 

A detailed table with all the significant connections for all the pairs is in the supplementary information (Table SI 4). In the figure, a panel with corresponding \textit{VI} measure value for the corresponding network pair for overall spectrum and the individual frequency bands are provided.

\begin{figure}[!h]
	\centering
	\includegraphics[scale=0.7]{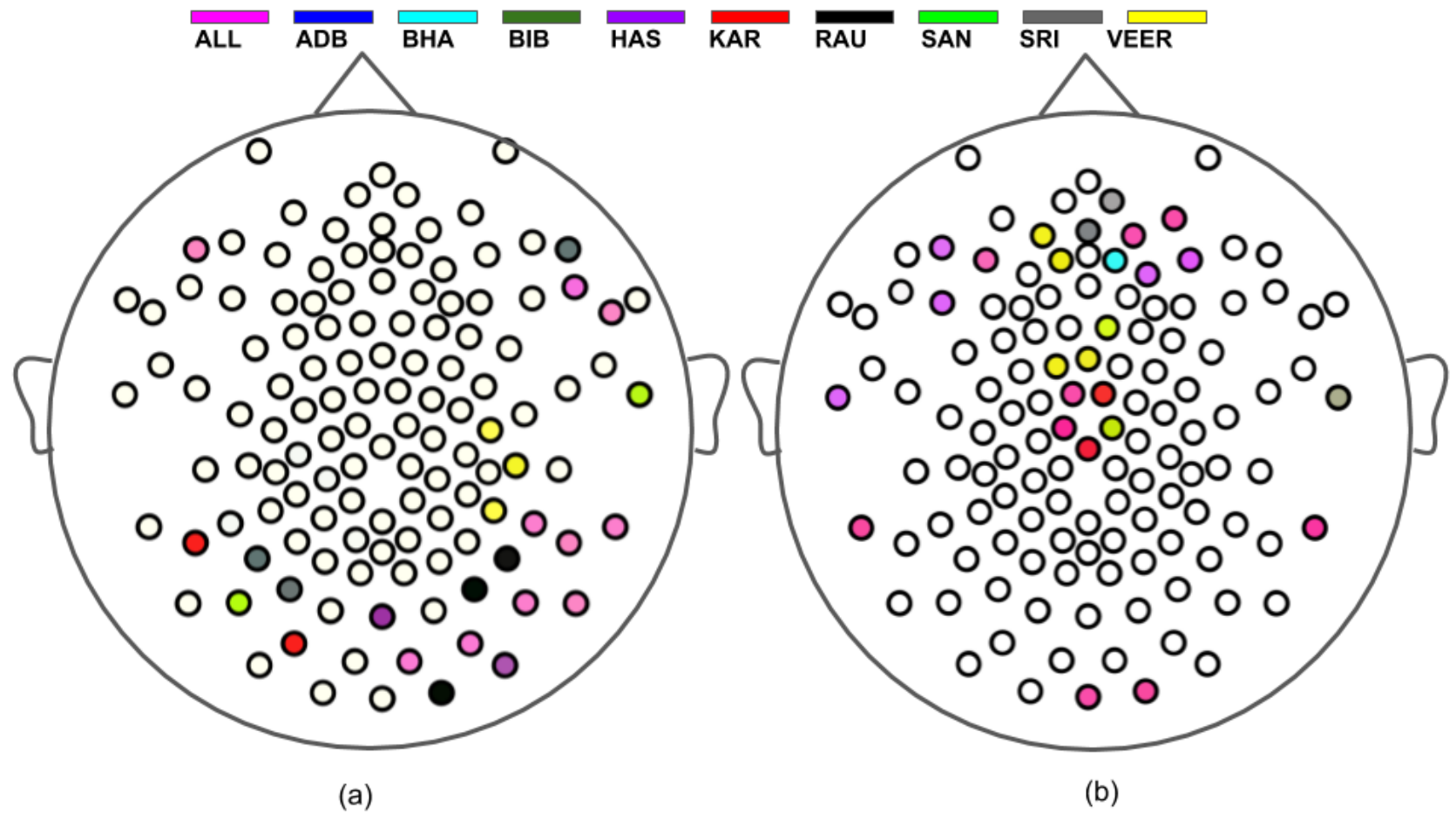}
	\caption{Brain map showing nodes with (a)highest positive leverage centralities (b) highest negative leverage centralities for all the \textit{Rasas}. Pink coloured nodes are the one which are consistently central for at least eight \textit{Rasas}, other colors are for \textit{Rasas} as shown in the color panel above.  }
	\label{fig: leverage}
\end{figure}

\subsection*{Hub Identification Using Centrality measure}
Hubs in the brain networks are nodes that are critical to information flow across the network and are identified based on various centrality measures such as Degree Centrality, Eigen-Vector Centrality\cite{newman2008mathematics}, Betweenness Centrality \cite{freeman1977set}, and Leverage Centrality\cite{joyce2010new}. As far as Brain networks are concerned, work by Joyce \textit{et al.} suggests that Leverage centrality \cite{joyce2010new} is computationally cheaper and more accurate for identifying hubs than other centrality measures, as revealed by their Receiver Operating Characteristic curves (see Methods for details). 

Leverage centrality also incorporates information about local node neighbourhood, such that a node with high positive leverage centrality is more impactful to its neighbours so that its neighbours draw more information from it than any other nodes in their neighbourhood. In contrast, a negative leverage centrality node is influenced more by its neighbours than being an influencer node. In the present study, we designate central nodes as nodes with the largest positive and negative leverage centralities.

In Fig.~\ref{fig: leverage}, are shown a brain map with coloured nodes having the highest (top 12) positive and highest negative (top 12) leverage centralities. Colour coding is based on the \textit{Rasas} as shown in the figure. Nodes shown in pink are consistently central across all the \textit{Rasas}.  For the positive centrality case, they all lie in the periphery of the brain map, mostly in the parietal brain regions indicating that most important information relay centres for the perception of emotions are located at the parietal regions of the brain. The central nodes with most negative leverage centralities are in 
the frontal and central brain regions demonstrating that nodes drawing maximum information from their neighbours are in these regions (In the Supplementary 
Information, we provide two tables (Table SI 2, Table SI 3) with labels of nodes with highest positive and negative leverage centralities for all the \textit{Rasas} ).

\section*{Discussion}

The \textit{Rasas} discussed in this work evoke emotions which are related to, but are different from the  western classification of emotions such as anger, fear, happiness, and sadness. But there is very little work done on the science of \textit{Rasas} and how they are perceived in the brain. To the best of our knowledge, this is the first attempt to design a network analysis probe to identify neural signatures for different \textit{Rasas} mentioned in Indian \textit{Natyasastra}. 

We have studied changes in the functional connectivity and modular network organisation in different frequency bands, concerning different \textit{Rasa} states, elicited in participants through watching of movie clips.  We have characterized how the emotional states are similar or dissimilar by measuring network segregation in different frequency bands. We find that the community structure in the alpha band is most similar across all \textit{Rasa} genres and low synchrony of nodes in higher frequency bands.  Previous studies have reported higher correlations for negative or stress full visual stimuli\cite{miskovic2010cross} for lower frequency bands. In line with this, our results indicate higher coherence in these bands for \textit{Raudra} and \textit{Bhayanaka} \textit{Rasas}.  

Our coherence analysis has also shown higher synchronization among nodes in lower frequency bands delta, theta and alpha and weaker correlations in beta and gamma bands for all the networks. We have identified the community structure of the brain for different \textit{Rasas} at the level of strongest links through which we can track the community evolution such as the birth, growth, merger and collapse of communities.   We find the existence of four dominant communities consistently across all \textit{Rasas} and localisation of hubs on brain map. We have also located the origins of coherence differences across \textit{Rasa} pairs in edges that are significantly different for these groups.  Finally, we identified most central nodes in all the networks using Leverage centrality and located them on the scalp, showing crucial centres for information processing. 

There are a few limitations to this work. Our observations are based on analysis of the signal space of electrodes placed on the scalp which does not have a clear source mapping inside the brain. Western classification of emotions such as those by Ekman, have been studied for decades using various imaging methods and analytical tools by researchers in various disciplines. However, we find very limited literature on quantitative study of \textit{Rasas}  based on brain imaging or EEG. Lack of reference limits standardization of our results. But we hope that our work will trigger interest in the research community to explore the scientific study of \textit{Rasas} further.

\section*{Materials and Methods}

Prior to conducting EEG experiments, ethical clearance was taken from the Institute Ethical Committee (IEC) of Indian Institute of Technology Gandhinagar.  Informed consent was obtained from all the participants before conducting experiments.

\subsection*{Subjects}
Participants were 20 healthy, right-handed students from Indian Institute of Technology Gandhinagar (mean age: 26 years, 16 males; 4 females). All of the participants were proficient in Hindi and English languages. They were all informed about the task and were asked to remain attentive while watching the film clip. We did an independent ranking of many movie clips corresponding to each of the \textit{Rasas} and only those
were selected which were ranked best suited to evoke a particular response among the \textit{Rasas}.

\subsection*{Movie Clips}
Complex, naturalistic stimuli like film-viewing evoke highly reliable brain activity across viewers as per current research works \cite{hasson2004intersubject}. In this work, we used 
nine film clips from Bollywood films (popular Indian Hindi language cinema) made between 1970's to the current time (see Table SI 1). These film clips were chosen to elicit the nine \textit{Rasas} among the viewers, based on the opinion of experts in \textit{Natyashastra}. Independent rating of the movies was done on a small number of subjects. The length of the movie clips varied from   42s to 2 mins 37s.  Based on \textit{Natyasastra} \cite{thirumalai2001introduction} human \textit{Rasa}s are classified into nine categories \cite{ekman1992argument} as given \textbf{in the background section} 

\subsection*{EEG Experimental Procedure}

We conducted the EEG experiment on the participants during which they were asked to watch chosen film clips representative of nine different \textit{Rasa} Genres. 
The electrical activity of the brain was recorded using 128 channel high-density Geodesic EEG Systems™ with a sampling frequency of 250Hz. A 
representative diagram of the node placement (along with node numbers from 1 to 128) used in the present work for brain network visualization, is shown in Fig.~\ref{fig:workflow} (a). The figure also depicts the electrode allocations to anatomical regions of the brain. The following abbreviations are used: F, L O and T stand for Frontal, Parietal, Occipital and Temporal lobes; L and R represent Left and Right regions. In Fig.~\ref{fig:com}, brain maps for the nine \textit{Rasa}s of the overall spectrum are shown. 


The participants were shown all the film clips in random order. A white fixation cross with a black screen for ten secs was shown before each clip. Initial few seconds of the time series recordings were neglected before analysis to avoid major fluctuations due to adjustment. Signals beyond 60 Hz frequencies were filtered out to avoid noise effects. The design of the experiment was done in E-prime™ and synced with Net-station™ acquisition software. The entire study on the collected data from the EEG experiment is summarised in the Fig.~\ref{fig:workflow}.

\begin{figure}[h]
	\centering    
	\begin{subfigure}{0.49\linewidth}
		\centering
	\includegraphics[width=0.70\linewidth]{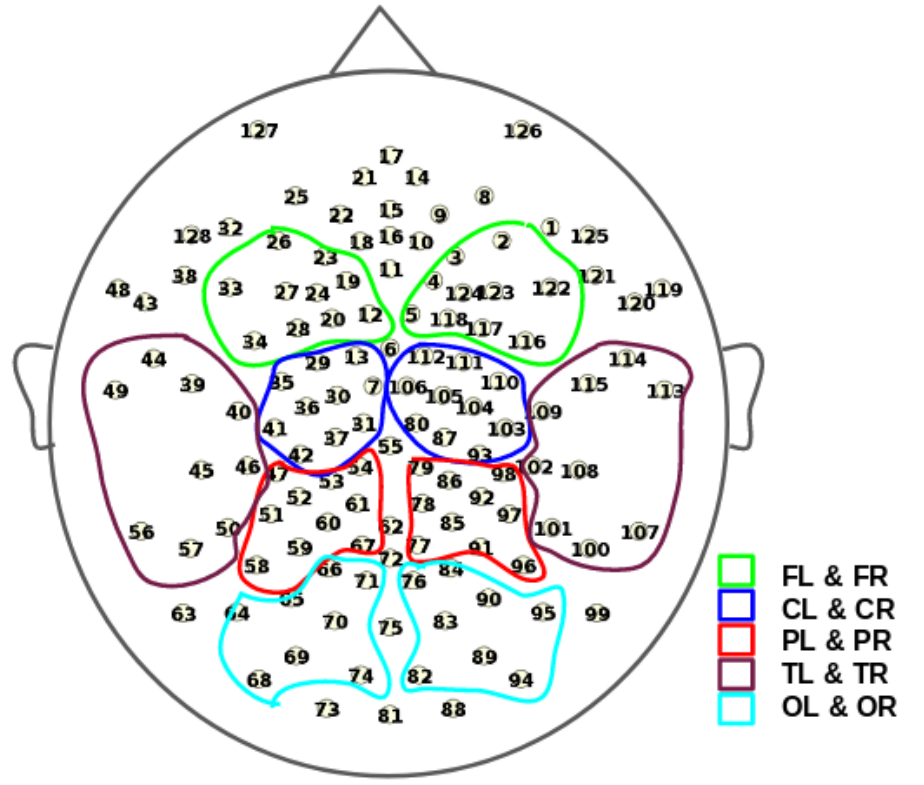}
	\caption{}
\end{subfigure}
	\begin{subfigure}{0.49\linewidth}
		\centering
	\includegraphics[width=1.0\linewidth]{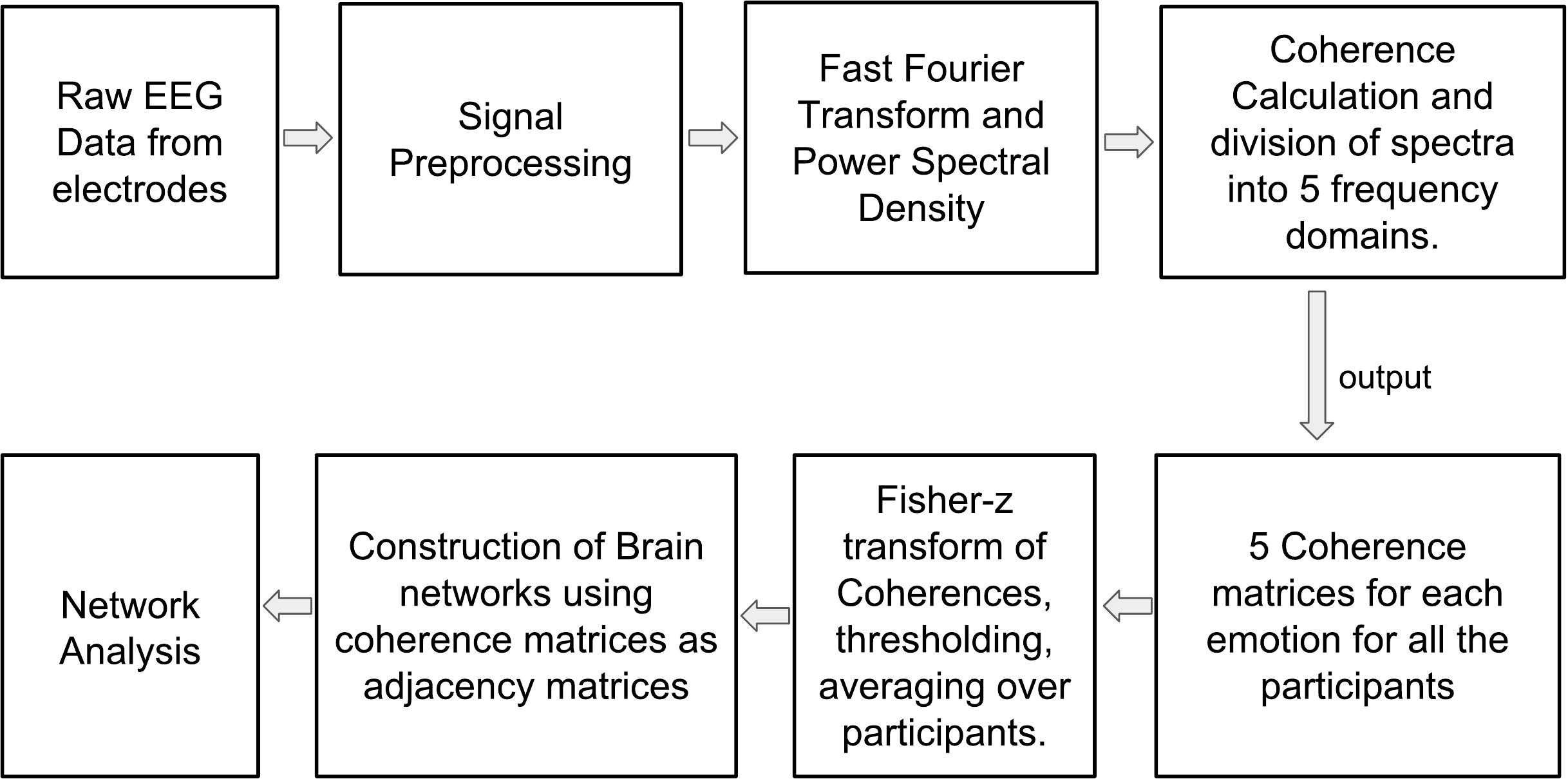}
	\caption{}
	\end{subfigure}
	\caption{(a) A brain map of the electrode positions of the EEG cap. All the Brain networks and extracted communities in the present work follow this Node placement. (b) Workflow structure governing EEG data recording, pre-processing, network construction and analysis}
	\label{fig:workflow}
\end{figure}

\subsection*{Construction of Brain Network}

Standard time series analysis measures such as Coherence \cite{bowyer2016coherence}, Phase Synchronization \cite{varela2001brainweb}, 
Mutual Information \cite{salvador2007frequency}, and Granger Causality \cite{hesse2003use} offer a huge depth into understanding the synchrony and 
information propagation in the functional brain. We use Electroencephalographic Coherence as a metric for deriving functional brain connectivity i. e. the degree of association between  any two brain regions (whose electrical activity is recorded by electrodes in the signal space). Holding values
between 0 (nil coherence) and 1 (complete coherence), it compares similarity of the power spectra (measured in microvolts squared, $\mu V^2$ ) of the time-series
recorded by these electrodes. The regions showing highest coherence are assumed to be the most synchronized functionally and vice-versa. The coherence measure (Eq.~\ref{eq:coherence}) between two time series $X$ and $Y$ is defined as,
\begin{equation}
C_{XY}(f) = \frac{|G_{XY}(f)|^2}{|G_{XX}(f)G_{YY}(f))|},
\label{eq:coherence}
\end{equation}
where $G_{XY}(f)$ is the cross-power spectral density and $G_{XX}(f)$ and $G_{YY}(f)$ are the respective auto-power spectral densities
\cite{thatcher1986cortico}.

The time series data (in microvolts, $\mu$V) was transformed to frequency domain via Fast Fourier Transform (FFT) and the power in different Brainwave 
frequency bands  \cite{tatum2014ellen}:  $\delta$ (1-4 Hz), $\theta$ (4 - 8 Hz), $\alpha$ (8 - 12 Hz), $\beta$ (12 - 40 Hz) and $\gamma$ (40 - 100 Hz), corresponding to different brain waves was calculated. The Coherence spectra, thus decomposed into five frequency bands resulted in five $128\times 128$ coherence matrices corresponding to each of the nine stimuli. The frequency decomposition of the spectrum was made to estimate the power in each of these bands and capture the brain state when a particular emotion was evoked. Apart from powers in individual bands we also used average coherence across the whole spectrum in the analysis. The coherence matrix was mapped to a weighted adjacency matrix, and corresponding brain network was constructed with  $N = 128$ nodes. In this network, the nodes are the electrode locations on the scalp and edges are connections between them measured by the coherence values.

Each of the electrodes from the EEG cap records the mean activity from a brain region and has contributions from the neighbouring brain areas as well, which may be spatially distant from the recording point. The resulting functional brain network is a dense network, retaining all the edges ($\mathcal{O}\sim N^{2}$) with non-zero edge weights. Also, in such functional networks, weak and non-significant links may represent spurious connections or false statistical dependencies. Such links tend to obscure the topology of stronger and significant connections and should ideally be discarded, by applying hard, or weighted thresholding. 
In the present work, the edges weights (coherences) representing functional correlations are Fischer Z-transformed, averaged over all the participants and then thresholded to retain only significant connections. To this end, 5 $\%$ of the weakest connections were omitted from the network, for all the nine cases. All the network analysis was done on these thresholded networks.

\subsection*{Network Analysis}

Network analysis of the functional networks performed in the present study consists of five elements as summarised in the Fig.~\ref{fig:Five_steps}. 
They are explained in the following sections.

\begin{figure}[htbp]
	\centering    
	\includegraphics[scale=0.6]{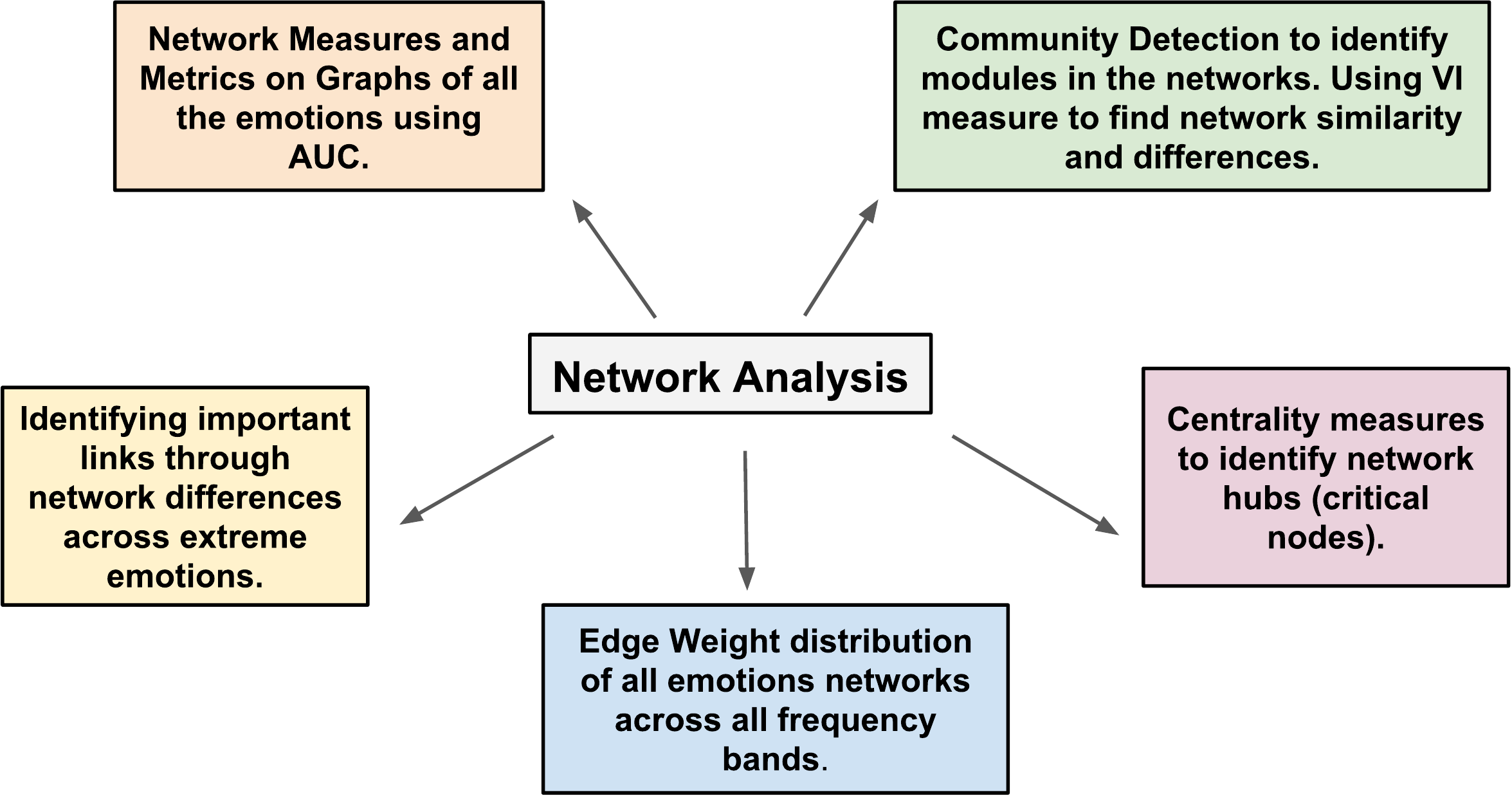}
	\caption{Block diagram of the methods used in the Network analysis of Brain Networks.}
	\label{fig:Five_steps}
\end{figure}

\subsubsection*{Small-worldness index} To distinguish between true functional connections from random connections among vertices, networks metrics must be compared to the corresponding mean values of a benchmark random graph with the same number of nodes and edges. We measured the small-worldness index of the networks defined as $[C/C_{rand}]/[L/L_{rand}]$, where $C_{rand}$ and $L_{rand}$ are the mean clustering coefficient and the characteristic path length of an ensemble of random networks with fixed number of nodes  $N$ and edges $\langle E\rangle= E_B$ of the Brain network. 

\subsubsection*{Area Under Curve method}
The network measures and metrics were evaluated on the un-thresholded networks, using the Area Under the Curve (AUC) approach\cite{bassett2006adaptive}. In the 
AUC procedure two cutoffs (one upper, one lower)on the edge strengths are consistently determined for each network, such that a range of important connections are retained, and only important network structure is dealt with.
The lower bound in this threshold range, $\kappa_{-}$, is an edge weight at which network is fully connected and the upper bound, 
$\kappa_{+}$ is the threshold at which the network just gets disconnected. The network is binarised at each threshold within this range, retaining only edges with weights larger than the threshold value and eliminating the remaining
ones and network metric ($M$) is evaluated at each connection density. This results in a curve showing the variation of $M$ with edge weights lying in the threshold range. The curve was then integrated over all these thresholds to yield the AUC value of the metric $M$.

\subsubsection*{Community detection and Graph Visualization}
For visualisation of dominant communities in the brain networks of all the \textit{Rasas}, we retained top 10 $\%$ of the edges and extracted communities on the obtained network. We used Gephi software\cite{bastian2009gephi} which utilises a modularity optimisation algorithm\cite{blondel2008fast} for detection of communities from the brain networks and Geolayout scheme for node positioning. The community structure of networks in different frequency bands for all the \textit{Rasas} is provided as animated graphic files in the supplementary material.

\subsubsection*{Leverage Centrality}
Leverage centrality  (defined in Eq~\ref{eq:LC}), 
\begin{equation}
\def\eqnwidth{\textwidth}
l_v = \frac{1}{k_v}\sum_{N_v}\frac{k_v - k_w}{k_v + k_w}
\label{eq:LC}
\end{equation}
measures the relationship between the degree of a vertex \textit{$k_v$} and the 
degree of each of its neighbors \textit{$k_w$}, averaging over all the neighbors $N_v$. For the weighted network, as in our case, degrees are weighted degrees.

\subsubsection*{Variation of Information}
Between two network partitions $X$ and $Y$, the $VI$ is defined in terms of the conditional entropy as below:
\begin{equation}
VI (X,Y)= H(X/Y) + H(Y/X).
\label{Eq:VI}
\end{equation}
The conditional entropy $H(X/Y)$ of partition $X$ given $Y$, with $M_x$ and $M_y$ modules can be computed as :
\begin{equation}
H(X/Y) = -\sum_{i=1}^{M_x}\sum_{j=1}^{M_y}p(x_i,y_i)\log\space p(x_i/y_j),
\label{Eq:cond_entropy}
\end{equation}
where $p(x_i, y_j) = n_{ij}/N$ is the joint probability of randomly selecting a node 
that belongs to modules $X_i$ and $Y_j$ and $p(X_i|Y_j) = n_{ij}/b_j$ is the conditional probability
that a node belongs to module $X_i$ in partition $X$, given that it is in module $Y_j$ in partition $Y$, and  $b_j$ is the number of nodes in community
$Y_j$ of partition $Y$. $n_{ij}$ is the number of nodes that are simultaneously present in module $X_i$  and $Y_j$.

\subsubsection*{Honest Significant Difference across \textit{Rasas}}

In our case, each of the \textit{Rasa}s was a group formed by twenty networks (corresponding to twenty participants coherence matrices) and mean of a particular edge weight was calculated over these twenty networks. The edge weight values in the coherence matrices corresponding to these groups were the ones tested for significance using the above two tests. So, practically we ran F-test $ 128\choose 2 $ (number of edges) times. The edges that were significant according to the F-test were subject to Tukey's test, to reveal which edge weight is significant across any two groups, with significance $\alpha \leq 0.05$. So, the two groups with significantly different mean were calculated for each edge weight, and in this group pair, one has a significantly higher or lower mean edge weight than the other.
These are depicted for six pairs in the Fig.~\ref{fig: HSD} in the results section. 

\section*{Data availability}
All the figures generated and used in the study, Gephi network files, coherence matrices, and tables  are uploaded on a Google drive folder (\href{https://drive.google.com/drive/folders/1E9--q3XJbM5iLd64o6RnUkIlAaodBWr4?usp=sharing}{Rasas$\_$BrainNetwork$\_$Data}). Raw EEG data can be provided upon request to authors. Details of the movie clippings (start and end times) are given in the Supplementary Information. 

\section*{Acknowledgments}
We would like to thank Dipanjan Roy from National Brain Research Center (NBRC), India and Pranjali Kulkarni, Indian Institute of Technology Gandhinagar (IITGn)  for insightful suggestions into this study. The study is a part of the Department of Science and Technology (India) sponsored Cognitive Science 
Research Initiative (DST-CSRI) project (Project number: SR/CSRI/PDF-08/2013, Dated 30/10/2014). RT would like to thank Amit Reza from IITGn for helpful discussions. We thank IITGn for providing access to the High-Performance Computing services.

\bibliography{references}

%
%

\end{document}